\documentclass[prd,twocolumn,showpacs,%preprintnumber,
floats,floatfix,
nofootinbib]{revtex4}
\usepackage{graphicx}
\usepackage{dcolumn}
\usepackage{bm}
\newcommand{\Hu}{{\cal H}} \newcommand{\Ka}{{\cal K}}
 
\newcommand{\dd}{{\rm d}}
\newcommand{\de}{{\rm e}}
\newcommand{\Pk}{{\Phi_k}\!\!}
\newcommand{\dphi}{{\delta\phi_k}\!\!}
\newcommand{\etai}{\eta_{\rm i}}
\newcommand{\etaj}{\eta_{\rm j}}
\newcommand{\etaB}{\eta_{_{\rm B}}}
\newcommand{\nS}{n_{_{\rm S}}}

\newcommand{\nT}{n_{_{\rm T}}}

\newcommand{\zN}{z_{_{\rm nucl}}}
\newcommand{\AS}{A_{_{\rm S}}}
\newcommand{\lP}{\ell_{_{\rm Pl}}}

\begin{document}

\author{Patrick Peter}
\email{peter@iap.fr}
\affiliation{Institut d'Astrophysique de Paris, UPR 341, CNRS, 98
boulevard Arago,  F-75014 Paris, France}
\author{Nelson Pinto-Neto}
\email{nelsonpn@cbpf.br}
\affiliation{Centro Brasileiro de Pesquisas F\'\i sicas,
Rue Dr.  Xavier Sigaud 150, Urca 22290-180 -- Rio de Janeiro, RJ,
Brazil}

\title{Primordial perturbations in a non singular bouncing universe
model.} 

\date{\today}

\begin{abstract}
We construct a simple non singular cosmological model in which the
currently observed expansion phase was preceded by a contraction. This
is achieved, in the framework of pure general relativity, by means of
a radiation fluid and a free scalar field having negative energy. We
calculate the power spectrum of the scalar perturbations that are
produced in such a bouncing model and find that, under the assumption
of initial vacuum state for the quantum field associated with the
hydrodynamical perturbation, this leads to a spectral index
$\nS=-1$. The matching conditions applying to this bouncing model are
derived and shown to be different from those in the case of a sharp
transition. We find that if our bounce transition can be smoothly
connected to a slowly contracting phase, then the resulting power
spectrum will be scale invariant.
\end{abstract}

\maketitle

%\pacs{Preprint number LPT-ORSAY 01-88}

%\narrowtext
%\vspace{0.2cm} ]

\section{Introduction}

For more than two decades, inflation~\cite{inflation} has been the
only available paradigm to solve the standard cosmological problems of
flatness, homogeneity, and monopole excess. It also predicts, as a
bonus, that primordial fluctuations, assumed to be of quantum origin,
could be enhanced to the level required to trigger large scale
structure formation, with an almost scale-invariant spectrum. To date,
no model has ever come close to challenging this impressive list of
successes.

Inflation cosmology suffers, however, from a few problems of its own,
whose seriousness is largely a matter of opinion. For instance, in a
typical realization, the underlying parameters (mass and coupling
constants of the inflaton field) must be assigned ``un-natural''
values in order to reproduce the observed temperature fluctuations in
the Cosmic Microwave Background Radiation (CMBR) which the mechanism
seeds. However, such a fine-tuning can be accounted for in various
realistic models.

The inflation paradigm is also endowed with two specific problems,
conceptually much more serious, that may ultimately be related, namely
the meaning of the trans-Planckian~\cite{transP} perturbations and the
existence of a past singularity~\cite{singularity}. Concerning the
latter, many ideas were discussed, among which the Tolman Ph\oe{}nix
universe~\cite{tolman}, and many others in the
seventies~\cite{seventies}, and recently revived under the name
``ekpyrotic''~\cite{ekp} in the somewhat different context of
superstring~\cite{strings} inspired brane cosmology~\cite{brane}.
This model, however, was the subject of many criticisms, both from the
string~\cite{noekp1} and cosmological~\cite{noekp2} points of view. In
its latest version~\cite{ekpN}, moreover, the model also contains a
supposedly actual singularity~\cite{mpps}.

Quantum cosmology, in the framework of the Wheeler-de-Witt (WdW)
equation, exhibits bouncing solutions~\cite{bounce} which can be
interpreted as truly avoiding the singularity, even for flat ($\Ka
=0$) spatial sections, a possibility strictly forbidden in classical
General Relativity (GR), unless~\cite{visser} some exotic material,
with negative energy density violating the Null Energy Condition
(NEC)~\cite{nec}, is introduced, which most cosmologists are reluctant
on doing. Such a bouncing universe model provides a solution to the
horizon problem by geodesically completing the manifold in the past,
and avoids the monopole formation if the bounce takes place at a
temperature below that of Grand Unification (GUT); this class of
models does not however address the question of flatness and one must
assume $\Ka=0$ from the outset. Moreover, in such a context, the
trans-Planckian issue simply does not exist because the initial
conditions for the perturbations can be imposed during a phase where
the universe is as close to the Minkowski spacetime as one wishes,
without ever passing through a Planck phase. Thus, it could be a
natural competitor to the inflationary paradigm, and it is therefore
of interest to estimate the primordial perturbation spectrum that it
can produce.

In a previous work~\cite{nobounce}, we examined the stability of a
bouncing universe dominated, at the bounce, by an exotic
hydrodynamical perfect fluid. We showed, by computing scalar
perturbations using the gauge invariant Bardeen~\cite{bardeen}
potential, that such perturbations grow unboundedly either at the
bouncing point, or at the time when the NEC was violated or restored,
thereby contradicting the hypothesis of low amplitude first order
perturbation theory~\cite{vectorial,mfb}. Such models are thus
incompatible with observational data, e.g., CMBR data~\cite{CMBdata}
according to which these first order effects indeed still dominate
(over nonlinear effects) at large scales.

The next to simple possibility consists on using a scalar field
instead of an exotic fluid. The purpose of this paper is to exhibit a
toy model in which a radiation fluid is coupled to a negative energy
free scalar field that is supposed to dominate the universe for a
limited amount of time, during which the bounce occurs~\cite{bfg}.
Our universe thus comes from a low-density, radiation dominated,
contracting state, passes through a bounce, and connects again back to
the usual Hot Big-Bang phase~\cite{SBB}.  In this context, we shall be
concerned with the scalar perturbations induced during the transition
between the collapsing and expanding phases.

In section~\ref{sec:model} we set the various constraints our model
needs to satisfy, and we explain how it can be made phenomenologically
reasonable. Then we calculate, in section~\ref{sec:pert}, the power
spectrum of the perturbations by matching the relevant solutions in
the various regions of interest, and we compare the results with
numerical calculations.  Contrary to what one would naively expect
from a fluid analysis~\cite{nobounce}, we find that scalar
perturbations are perfectly well behaved all along.

Setting vacuum initial conditions for the quantized hydrodynamical
perturbations deep in the low-density radiation dominated phase, we
find that the relevant spectrum of perturbations, at last horizon
crossing during the expanding radiation dominated era, has a spectral
index $\nS=-1$. It is thus incompatible with observational data. As it
is a model dependent result, further investigations of more realistic
models~\cite{bounce2,PBB}, from the point of view of particle physics,
need to be done~\cite{fppn}.

While other models yield a scale invariant spectrum by making use of
various assumptions~\cite{bf2}, the present calculations are made with
a specific model where the transition through the bounce is made with
an exact solution. This allows one to obtain, qualitatively and
numerically, the transitions in the Bardeen potential and its
derivative through the bounce, yielding indications on what kind of
matching conditions~\cite{matching} should be proposed for
perturbations passing through a general bounce. We obtain the perhaps
not so surprising result that the Bardeen potential changes sign
through the bounce, even though its derivative is continuous, contrary
to the case of a sharp transition~\cite{conserved,Ruth}. This result
will be discussed in more detail in section~\ref{sec:conclusion}, in
which we present a way to obtain a scale invariant spectrum for the
scalar perturbation by connecting our bounce and radiation dominated
model with a slowly contracting phase. If such a four-dimensional and
singularity-free model can be constructed, it will be able to
reproduce all the available observational data while avoiding most of
the questions raised by the inflation solution.

This paper only deals with the scalar part of the perturbations, which
we show does not yield a spectrum compatible with the data.  The
tensor part was already calculated in Ref.~\cite{mpps}, where the
spectral index $\nT = \nS -1=2$ was obtained, and is therefore unable
to reproduce the data. Finally, we will not be concerned with
vectorial (rotational) perturbations even though, contrary to the
usual inflationary case, one could think that those have no reason to
be {\sl a priori} negligible with a time symmetric scale
factor. However, the Universe is torque-free~\cite{grishchuk} since at
least the nucleosynthesis epoch that occurred at a redshift of
$\zN\sim 3\times 10^8$. Hence, the present relative contribution
$\delta_v$ for the vectorial perturbation, which scales as
$a^{-2}$~\cite{vectorial}, is expected to be $\delta_v\ll 10^{-17}$,
independent of the scale $k$ at which it is evaluated, and hence
observationally irrelevant.

\section{The model}
\label{sec:model}

We shall consider a very simple toy model for which we demand the
following conditions to hold. First of all, we want general
relativity to be valid for all times. We also impose that at late
times, the model should reproduce the standard hot big bang case,
i.e. there should exist a time in which radiation dominates. This
implies in particular that we assume some amount of radiation to
be present in our model. We also restrict our attention to the
spatially flat situation. Finally, the model should have a
bouncing phase. This means, given that there is already some
radiation present, that, in the context of GR, there must exist
some other fluid having negative energy. In particular, for the
special case at hand for which the spatial curvature $\Ka=0$,
this means that the Null Energy Condition (NEC) must be violated
at some time near the bounce~\cite{visser}.

Realizing such a model is in principle feasible with just another
fluid, e.g., some stiff matter with negative energy, namely one for
which the equation of state reads $p=\rho<0$. However, it was recently
shown~\cite{nobounce} that such an approach will lead to an
overproduction of large inhomogeneities at various different times,
breaking the cosmological principle hypothesis long before
nucleosynthesis. Such an approach is therefore not applicable, and we
must resort to the next-to-simple possibility, namely a free massless
scalar field which is known to reproduce the stiff matter fluid
behavior at the background level~\cite{bfg}. The action we shall start
with thus reads \begin{equation} {\cal S} = \int \left( -{1\over 16\pi
G} R - \epsilon - {1\over 2} \nabla_\mu \phi \nabla^\mu \phi \right)
\sqrt{-g}\, \dd^4x ,\label{action} \end{equation} where $R$ is the
curvature scalar, $\epsilon$ the energy density of the radiation
fluid, and $\phi$ the scalar field. We assume that the background
metric takes the standard Friedmann-Robertson-Walker form
\begin{equation} \dd s^2 = a^2(\eta) \left( \dd\eta^2 - \delta_{ij}
\dd x^i \dd x^j \right),
\label{FRW}\end{equation}

\noindent with $\eta$ the conformal time. The cosmic time, $t$, is
then obtained as the solution of the equation $a \dd\eta = \dd t$ once
the scale factor $a(\eta)$ is known. Note that throughout this paper,
we assume the background curvature to vanish, $\Ka=0$. In this
context, it has to be a particular choice: this category of models
does not indeed solve the flatness problem.

Varying the action~(\ref{action}) with respect to the fluid and fields
yields the background dynamical equations
\begin{equation} \varepsilon' + 4\Hu\varepsilon = 0, \ \
\varphi''+2\Hu \varphi'=0,
\label{back} \end{equation}

\noindent where $\Hu\equiv a'/a$, $\varepsilon$ and $\varphi$ are the
background space-independent values of the radiation energy density
$\epsilon$ and scalar field $\phi$ respectively, and a prime denotes a
differentiation with respect to the conformal time $\eta$. These
background equations imply \begin{equation} \varphi' = {c\over a^2}, \
\ \ \varepsilon = {d\over a^4},\label{cd}\end{equation} where $c$ and
$d$ are constant. The energy density of the scalar field is given by
\begin{equation}
\rho _{\varphi} \equiv -{\varphi^{\prime 2}\over 2a^2} =-{c^2\over 2a^6},
\end{equation}
and as such it is dominant when $a$ is small and negligible when 
$a$ is very large.
These solutions, together with Friedmann equation
\begin{equation} \Hu^2 = \lP^2 \left( a^2 \varepsilon -{1\over 2}
\varphi^{\prime 2}\right), \ \ \ \lP^2 \equiv {8\pi G\over
3},\label{fried}
\end{equation} \noindent lead to the bouncing solution \begin{equation}a(\eta) =
a_0 \sqrt{1+\left({\eta\over \eta_0}\right)^2},
\label{scale} \end{equation}

\noindent where the minimum scale factor $a_0$ and the characteristic
bouncing conformal time $\eta_0$ solely depend on the relative
quantities of energy density in radiation and scalar field at some
given time: $a_0^2 = c^2 / (2 d)$ and $\eta_0^2 = c^2/(2 d^2 \lP^2)$. In
what follows, these two parameters will be considered as the relevant
ones.

\begin{figure}[t]
%\begin{center}
\includegraphics*[width=8.5cm]{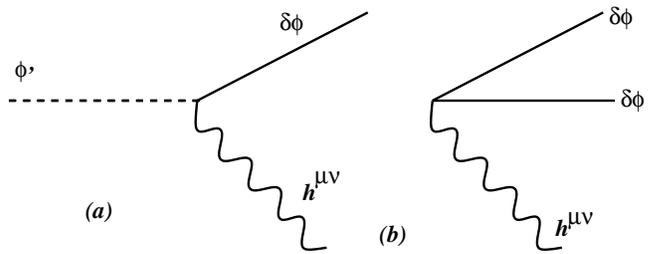}
%\epsfig{file=feynman.eps,width=8cm}
%\vskip6mm
\caption{Diagrams leading to instabilities in the
theory~(\ref{action}). $(a)$: the dynamical instability whereby
the energy contained in the scalar field can be used to produce
semi-classical perturbations, later to be identified with
primordial fluctuations. $(b)$: Vacuum instability. As this
process in non zero, the vacuum can spontaneously decay into a
pair of negative energy scalar particles and a positive energy
graviton.} \label{feynman}
%\end{center}
\end{figure}

Before turning to first order perturbations of this background, which
is the subject of the following section, we want to emphasize a point
of stability of this model related to the ``wrong'' sign chosen in
Eq.~(\ref{action}). Indeed, an expansion of Eq.~(\ref{action}) with
respect to
\begin{equation} g_{\mu\nu} = g^{^{\rm (0)}}_{\mu\nu} + h_{\mu\nu}, \
\ \ \phi = \varphi +\delta \phi ,\end{equation}

\noindent with $[{\mathbf g}^{^{\rm (0)}},\varphi]$ the classical
part and $({\mathbf h},\delta\phi)$ interpreted respectively as
gravitons and scalar particles in a semi-classical approach, will
inevitably lead to two different kinds of instabilities, each
arising at a different order in perturbation. The first one, with
which we shall deal later since it is actually the one
responsible for the large scale structure formation in this
model, is second order in perturbation (first order in the
equations of motion) and goes essentially as $\propto h^{\mu\nu}
\partial_\mu\delta\phi\partial_\nu\varphi$. This term is absent
in ordinary Minkowski space, but is present in the cosmological
setup we are considering because of Eq.~(\ref{back}) in which the
classical part of the scalar field varies with time and thus
behaves as a source for the production of gravitons and scalar
particles. As it originates in a derivative coupling [see
FIG.~\ref{feynman}$-(a)$], the characteristic time scale of this
instability is that of the classical scalar part, in our case the
typical cosmological timescale.

The second instability that must be discussed is much more
serious, even though, at first sight, it looks innocuous because
of a higher order in perturbation: it is the same term as before,
but with the classical part replaced by a first order
perturbation, namely $\propto h^{\mu\nu}
\partial_\mu\delta\phi\partial_\nu\delta\phi$ [see
FIG.~\ref{feynman}$-(b)$]. The presence of such a process means
that the vacuum can spontaneously decay into a pair of negative
energy scalar particles and a graviton, and, due to this fact,
the energy levels are not bounded from below. This sounds like a
catastrophe, and even more so because the only available timescale
comes from the coupling constant, i.e., the Planck time. However,
it is clear from the figure that the process probability
amplitude ${\cal A}$ is ${\cal A}\propto p^2 /M_{_P}^2$, with
$M_{_P}\sim \lP^{-1} \simeq 10^{19}\,$GeV the Planck mass and $p$
the momentum at the vertex. Such an amplitude therefore becomes
important when the characteristic scale $p^{-1}$ is comparable to
$\lP$. At this point, it should be argued that the model of
Eq.~(\ref{action}) is understood as an effective low energy
theory which must be implemented with a cutoff scale much larger
than the Planck one: as one reaches the Planck energy scale, the
theory is expected to break down into a completely different one
such as, e.g., quantum gravity or superstring theory. As a result,
for cosmological purposes, one can safely ignore this instability
and concentrate on the production of cosmological perturbations.

\hskip3mm
\section{Linear perturbation spectrum}
\label{sec:pert}

In what follows, we shall consider perturbations stemming from
the model~(\ref{action}), making use of the gauge invariant
formalism~\cite{bardeen,mfb}. As there are no anisotropic stress
perturbations in this model, the most general form of metric
perturbations on the background given by Eq.~(\ref{FRW}) reads, in
the longitudinal gauge,

\begin{equation} \dd s^2 = a^2(\eta) \left[ (1+2 \Phi) \dd\eta^2 -
(1-2\Phi)\delta_{ij} \dd x^i \dd x^j \right],\label{dg}\end{equation}

\noindent where $\Phi$ is the gauge invariant Bardeen
potential~\cite{bardeen}. Setting also

\begin{equation} \phi = \varphi (\eta) + \delta\phi ({\mathbf x},\eta)
\ \ \ \hbox{\rm and} \ \ \ \epsilon = \varepsilon (\eta) + \delta
\epsilon ({\mathbf x}, \eta),\label{dphieps}\end{equation}

\noindent one obtains the radiation fluid current conservation and
Klein-Gordon equation respectively in the form

\begin{equation}
\left\{
\matrix{\delta\epsilon'+4\Hu \delta\epsilon = {4\over 3}
\varepsilon \left( 3 \Phi' + a^{-1} \nabla^2 \tilde\alpha\right) , \hfill\cr
\hfill\cr \delta\phi''+2\Hu \delta
\phi'-\nabla^2\delta\phi=4\Phi'\varphi',\hfill } \right.
\label{dmat}
\end{equation}

%\begin{eqnarray}
%\delta\epsilon'+4\Hu \delta\epsilon &=& {4\over 3}
%\varepsilon \left( 3 \Phi' + a^{-1} \nabla^2
%\tilde\alpha\right),\nonumber  \\
%\delta\varphi''+2\Hu \delta \phi'-\nabla^2\delta\phi&=&4\Phi'\varphi'
%%\right.
%\label{dmat}
%\end{eqnarray}

\noindent where $\tilde\alpha$ is the gauge invariant fluid velocity
potential, and use has been made of the relation $\delta\epsilon =
{1\over 3} \delta p$ between the energy density $\delta\epsilon$ and
pressure $\delta p$ fluctuation. Einstein equations yield, after a bit
of algebra~\cite{mfb},
%\begin{widetext}

%\begin{equation} \left\{ \matrix{ \Phi'+\Hu\Phi={3\over 2} \lP^2 \left(-
%\varphi'\delta\phi +{4\over 3} a \tilde\alpha\right) ,\hfill&\cr
%\hfill&\hfill\cr \nabla^2\Phi -3\Hu\Phi'-3\Hu^2\Phi = {3\over 2}\lP^2
%\left( -\varphi'\delta\phi' +\varphi'^2\Phi+a^2\delta\epsilon\right),
%\hfill&\cr \hfill\cr \Phi''+3\Hu\Phi'+\left(2\Hu'+\Hu^2\right) \Phi =
%{3\over 2}\lP^2 \left( -\varphi'\delta\phi'+\varphi'^2\Phi+{1\over
%3}a^2\delta\epsilon\right).&}\right.  \label{dG}\end{equation}

\begin{equation} \left\{ \matrix{ \Phi'+\Hu\Phi={3\over 2} \lP^2 \left(-
\varphi'\delta\phi +{4\over 3} a \tilde\alpha\right) ,\hfill&\cr
\hfill&\hfill\cr \nabla^2\Phi -3\Hu\Phi'-3\Hu^2\Phi = {3\over 2}\lP^2
\left( -\varphi'\delta\phi' +\varphi'^2\Phi+a^2\delta\epsilon\right),
\hfill&\cr \hfill\cr \Phi''+3\Hu\Phi'+\left(2\Hu'+\Hu^2\right) \Phi =
{3\over 2}\lP^2 \hfill&\cr
\hfill\times\left( -\varphi'\delta\phi'+\varphi'^2\Phi+{1\over
3}a^2\delta\epsilon\right).&}\right.  \label{dG}\end{equation}

%\begin{eqnarray} \Phi'+\Hu\Phi&=&{3\over 2} \lP^2 \left(-
%\varphi'\delta\phi +{4\over 3} a \tilde\alpha\right) ,\nonumber \\
%\nabla^2\Phi -3\Hu\Phi'-3\Hu^2\Phi &=& {3\over 2}\lP^2 \left(
%-\varphi'\delta\phi' +\varphi^{\prime 2}\Phi+a^2\delta\epsilon\right),
%\nonumber \\ \Phi''+3\Hu\Phi'+\left(2\Hu'+\Hu^2\right) \Phi &=&
%{3\over 2}\lP^2 \left( -\varphi'\delta\phi'+\varphi^{\prime
%2}\Phi+{1\over 3}a^2\delta\epsilon\right).  \label{dG}\end{eqnarray}
%\end{widetext}
\noindent Simple manipulations of Eqs.~(\ref{dmat}) and (\ref{dG})
permit to eliminate the radiation fluctuation in favor of the Bardeen
potential through the relation
\begin{equation}
\Pk''+6\Hu\Pk'+\left[2\left(\Hu'+2\Hu^2\right)+k^2\right] \Pk = -\lP^2
a^2 \delta\epsilon_k,\label{deps}\end{equation}

\noindent where, from now on, we assume a Fourier decomposition of
each variable $A$ into its components $A_k$ defined through
\begin{equation} A_k (\eta) \equiv \int {\dd^3 x\over
(2\pi)^{3/2}}\de^{-i{\mathbf k}\cdot {\mathbf x}} A(|{\mathbf
x}|,\eta),\end{equation} where $A_k$ only depends on the amplitude $k$
of the wave-vector ${\mathbf k}$.

The dynamical equations for the Bardeen potential and the fluctuations
of the scalar field therefore decouple from the radiation fluid
perturbations and are then expressible solely in terms of themselves
as (making use of the background Einstein equations) \begin{equation}
\Pk''+4\Hu\Pk'+{1\over 3} k^2 \Pk = -\lP^2 \varphi' \dphi',
\label{Bardeen}\end{equation} and
\begin{equation}\dphi''+2\Hu\dphi'+k^2\dphi = 4 
\varphi'\Pk'. \label{scalar}\end{equation} We shall now investigate
the solution of these equations in order to get the perturbation
spectrum such a bouncing model predicts.

\subsection{The relevant phases in the perturbations evolution}

\begin{figure}[t]
\includegraphics*[width=8.5cm]{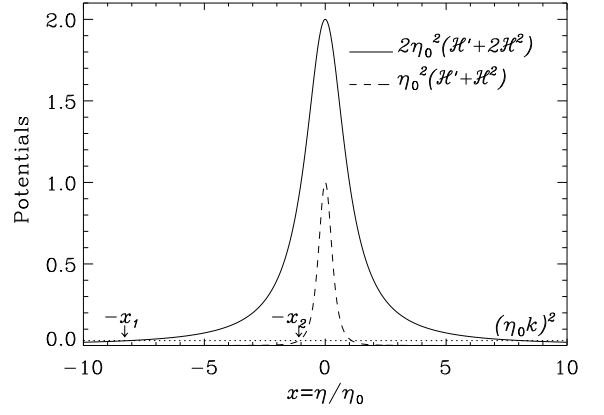}
\caption{Potentials for the parametric oscillator equations giving the
dynamics of the Bardeen potential and the scalar field
perturbations. The full line shows the potential for the variable
associated with $u_k$ [see Eq.~(\ref{uk})], the dashed line the
potential for the scalar field $w_k$ [Eq.~(\ref{wk})], and the dotted
line, showing the value of $(\eta_0 k)^2$, indicates visually the
different regions where the different approximations hold. The points
$x_1$ and $x_2$ are the matching points for these two fields.}
\label{fig:pots}
\end{figure}

In order to investigate Eqs.~(\ref{Bardeen}) and (\ref{scalar}), let
us write them in terms of the variables $u_k\equiv a^2\Pk$, and
$w_k\equiv a\dphi$. Using Eq.~(\ref{cd}), they read
\begin{equation}
u_k''+\left[{1\over 3} k^2-2\left(\Hu'+2\Hu^2\right)\right]u_k =
-\lP^2 {c\over a}\left(w_k' -\Hu w_k\right),
\label{Bardeen2}\end{equation} and
\begin{equation} w_k''+\left(k^2-\Hu
^2-\Hu '\right)w_k = 4{c \over a^3}\left(u'_k-\Hu u_k\right).
\label{scalar2}
\end{equation}
Each of these equations can be seen as an inhomogeneous equation,
i.e., one with a source term not depending on the function
itself. Asymptotically, since the scale factor grows like $|\eta|$, it
can be checked explicitly, by means of an expansion in powers of
$\eta^{-1}$ for $u_k$ and $w_k$, that the source terms in both
Eqs.~(\ref{Bardeen2}) and (\ref{scalar2}) are small for\footnote{This
is valid not only when $\eta\to\infty$ but also when $k^2<6(\Hu'+2\Hu
^2)$ as long as $\eta\gg\eta_0$. We shall return to this point in more
detail below.}  $\eta\gg\eta_0$.  Therefore, in this limit,
Eqs.~(\ref{Bardeen}) and (\ref{scalar}) can be reduced to the usual
parametric oscillator equations for the variables $u_k$ and $w_k$,
namely
\begin{equation} u_k'' + \left[{1\over 3}k^2 - {(a^2)''\over a^2}
\right] u_k =0,\label{uk}\end{equation}
and
\begin{equation} w_k'' + \left(k^2 - {a''\over a}
\right) w_k =0.\label{wk}\end{equation} The potentials $(a^2)''/ a^2
= 2(\Hu'+2\Hu^2)=2/(\eta^2+\eta_0^2)$ for $u_k$ and 
$a''/a=\Hu'+\Hu^2=\eta_0^2/(\eta^2+\eta_0^2)^2$ for $w_k$ are shown on
FIG.~\ref{fig:pots} on which we also define the variable
$x\equiv\eta/\eta_0$, as well as the various corresponding matching
values $x_1$ and $x_2$.

We shall be interested in the cosmologically relevant limit $k\ll 1$.
However, as $|\eta|\to\infty$, the $k$-dependent terms in
Eqs.~(\ref{uk}) and (\ref{wk}) become important. More precisely, when
$k|\eta|>\sqrt{6}$, i.e., when $|x|> x_1 \equiv\sqrt{6}/(k\eta_0)
\gg 1$, the solutions of the above equations in terms of $\Pk$ and
$\dphi$ can be written in terms of Hankel functions,
namely~\cite{Grad},

\begin{equation} \left\{ \matrix{
\Pk^{^{\rm rad}} = \eta^{-3/2} \left[ \Phi_{_{(1)}} H^{^{\rm
(1)}}_{3/2} (\omega\eta) + \Phi_{_{(2)}} H^{^{\rm (2)}}_{3/2}
(\omega\eta) \right], \hfill &\cr \hfill &\cr \dphi^{^{\rm rad}} =
\eta^{-1/2} \left[ X_{_{(1)}} H^{^{\rm (1)}}_{1/2} (k \eta) +
X_{_{(2)}} H^{^{\rm (2)}}_{1/2} (k \eta) \right],\hfill&}\right.
\label{radiation}\end{equation}
where $\omega=k/\sqrt{3}$.

When the potential terms dominate over the $k$-dependent terms, which
for $u_k$ is the case so long as $k|\eta| \ll\sqrt{6}$, or $1\ll
|x|\ll x_1$, and for $v_k$ when $k|\eta| \ll \sqrt{k\eta_0}$, or $1\ll
|x|\ll x_2$, where $x_2\equiv 1/\sqrt{k\eta_0}$, the zeroth order
solutions for Eqs.~(\ref{uk}) and (\ref{wk}) read, when $x_1\ll x\ll
x_2$,

\begin{equation} 
\Pk^{<}=A_1 + A_2\int a^{-2}\dd\eta + O(k\eta) 
\approx A_1-A_2{\eta_0^4\over 3a_0^4\eta ^3},
\label{phi2}
\end{equation}

\begin{equation} 
\dphi^<=B_1 + B_2\int a^{-1}\dd\eta + O(k\eta)
\approx B_1-B_2{\eta_0^2\over a_0^2\eta}.
\label{sca3}
\end{equation}
In fact, solution (\ref{sca3}) needs amelioration because around
$|x|=x_2$ the source term of Eq.~(\ref{scalar}) become important. This
is not the case of solution (\ref{phi2}) because for $1\ll |x|<x_2$,
the source terms of Eq.~(\ref{Bardeen}) are still negligible, even
taking into account the corrections of (\ref{sca3}). Fortunately, for
what follows in the next subsection, only the solution (\ref{phi2})
will be needed.

\begin{figure}[t]
\includegraphics*[width=8.5cm]{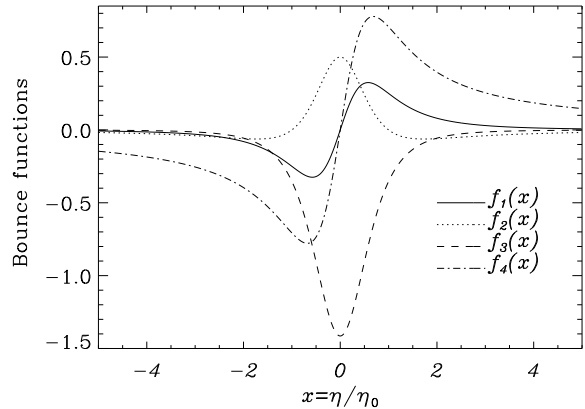}
\caption{The bounce functions $f_i$ as functions of $x\equiv\eta/\eta_0$.}
\label{fig:fi}
\end{figure}

Near the bounce, when the potentials and $c/a$ are of order
$1/\eta_0^2$ and $a_0/(\eta_0\lP)$, the source terms become
important\footnote{The constraints on the values of $a_0$ and $\eta_0$
in order for the bounce to happen at a scale much larger than the
Planck scale and a long time before nucleosynthesis takes place are
$\lP\ll a_0\eta_0\ll 10^8{\rm cm}$. Hence, if one chooses $\eta_0$ of
order one, the potentials are of order one, and $c/a\gg1$ near the
bounce.}, but the terms proportional to $k^2$ are still negligible.
In this situation, one can neglect altogether the $k^2$ term in
Eqs.~(\ref{Bardeen}) and (\ref{scalar}), yielding the solutions
\begin{equation} \Phi^{^{\rm Bounce}} = \tilde{A} + \tilde{B} f_1(x)
+\tilde{C} f_2(x),\label{Phi0}\end{equation}
\noindent and
\begin{equation} \lP\delta\phi^{^{\rm Bounce}} = \tilde{D} + \tilde{B}
f_3(x)+ \tilde{C} f_4(x), \label{X0}\end{equation}
\noindent with $\tilde{A}$, $\tilde{B}$, $\tilde{C}$ and $\tilde{D}$
arbitrary constants. The bounce functions $f_i(x)$ are found to be
\begin{equation} f_1 (x) \equiv {x\over (1+x^2)^2},\ \ \ \ f_2(x)
\equiv {1-x^2 \over 2(1+x^2)^2},\end{equation}
\begin{equation} f_3 (x) \equiv -{\sqrt{2}\over (1+x^2)^2},\ \ \ \ f_4(x)
\equiv {x\over\sqrt{2}} {3+x^2 \over (1+x^2)^2},\end{equation}
and are displayed on FIG.~\ref{fig:fi}.

These solutions will be used to match the asymptotic solutions through
the bounce.

\subsection{Matching the solutions, and the power spectrum}

In the limit $\eta\to\pm\infty$ ($\Longleftrightarrow a\to\pm\infty$),
i.e., very far from the bounce, the Universe is radiation dominated,
so that the coupling term in the left hand side of
Eqs.~(\ref{Bardeen2}) and (\ref{scalar2}) can be neglected, as it was
explained in the last subsection. From Eq.(\ref{radiation}), and for
$\eta\to -\infty$, the Bardeen potential and the scalar field
perturbation respectively scale as $1/a^2$ and $1/a$. From
Eq.~(\ref{deps}) one can see that the fluid perturbation $\delta
\epsilon _k$ goes like $1/a^4$.  Therefore, from Eq.~(\ref{dG}), we
find that the scalar field and its perturbation are irrelevant, in
this regime, for the evolution of the Bardeen potential with respect
to the radiation fluid.  For this reason, one can conclude that the
appropriate quantum gauge invariant variable to be used must be the
same as the one defined in Ref.~\cite{mfb} for the quantum treatment
of hydrodynamical fluids perturbation theory, which, in the case of
pure radiation, is related to $\Phi$ by,
\begin{equation} 
\Phi _k = 3\sqrt{{3\over 2}}{\lP\beta\over \Hu k^2}
\biggl({v_k\over z}\biggr)^{\prime},
\label{v}
\end{equation}
where $\beta = \Hu^2 - \Hu '$ and $z\equiv a\sqrt{3\beta}/\Hu$.
Similarly, the gauge invariant quantum variable connected to the
scalar field perturbation given in Ref.~\cite{mfb} is given by
\begin{equation} 
\omega _k = a[\delta\phi + (\varphi/\Hu)'\Phi] \approx w_k.
\label{w}
\end{equation}
It is interesting to note that the quantum field $v$ leaves the
oscillatory regime at the same conformal time as $\delta\phi$ does,
and that neither of them do so at horizon crossing. This is a
peculiarity of our model due to the fact that at the time at which the
quantum fields leave the oscillatory regime, the space is almost
radiation dominated, but not quite.

Imposing the initial vacuum state for these quantum variables implies
that we can set
$$v_k = {3^{1/4}\de^{-ik\left(\eta-\etai\right)/\sqrt{3}} \over
\sqrt{2k}}$$
and
$$w_k = {\de^{-ik\left(\eta-\etaj\right)}\over \sqrt{2k}}$$ at
$\eta\to -\infty$, with $\etai$ and $\etaj$ two {\sl a priori}
arbitrary conformal times, having no influence on the subsequent
evolution. From the solutions (\ref{radiation}) and these initial
conditions, one can write the Bardeen potential $\Pk$ and the scalar
field perturbation $\dphi$ at $k\eta \ll -\sqrt{6}$ (or $|x|\gg
x_1\gg 1$) as

\begin{equation} 
\Pk^{^{\rm ini}} = -{\lP\eta _0 3^{3/4}\over 2a_0\eta ^2 k^{3/2}}
\biggl({\sqrt{3}\over k\eta}+i\biggr)\de^{-ik(\eta-\etai)/\sqrt{3}},
\label{phii}
\end{equation}
and
\begin{equation} 
\dphi^{^{\rm ini}} = {\eta _0\over a_0\eta\sqrt{2k}}
\de^{-ik(\eta-\etaj)}.
\label{scai}
\end{equation} 

We are interested in calculating the power spectrum

\begin{equation}
{\cal P}_k\equiv k^3 |\Pk\,\,|^2 \equiv \AS k^{\nS-1},
\label{spectrum}
\end{equation}
evaluated at the time when $\Pk$ returns to its oscillatory regime,
i.e., at $x=x_1$. As we shall see later, the values of $\dphi$ in the
different phases of perturbation evolution are not necessary to
calculate $\Pk$ at $x=x_1$.  Hence, we will forget about $\dphi$ from
now on.

Looking at Eq.~(\ref{Bardeen2}), one can see that the first matching
must be imposed when $k^2/3 = 2(\Hu'+2\Hu^2) = (a^2)''/a^2$, for $u_k
= a^2 \Pk$. As $k$ is very small, this happens when $|\eta|\gg 1$
(where we can ignore the source terms).  Matching the solution
(\ref{phii}) with solution (\ref{phi2}) at the point
$k\eta\approx-\sqrt{6}$ (or $x\approx -x_1$) yields
\begin{equation} 
A_1 = {\lP\eta _0\sqrt{k}\over 3^{5/4}2\sqrt{2}}\de^{i\left(\sqrt{2}+
k\etai\right)/\sqrt{3}},
\label{A1}
\end{equation}
and
\begin{equation} 
A_2 = {\lP a_0^3 3^{5/4}\over 2\eta _0^3 k^{5/2}}(1-3\sqrt{2}i)
\de^{i\left(\sqrt{2}+k\etaj\right)/\sqrt{3}}.
\label{A2}
\end{equation} 

The solution (\ref{phi2}) is valid up to the point where $x$ is of
order one, when we approach the bounce. Differentiating
Eq.~(\ref{Bardeen}) twice and making use of Eq.~(\ref{scalar}) as well
as the background equations, we obtain the following fourth order
equation
\begin{widetext}
\begin{equation} 
\Pk^{^{\rm (IV)}}+10\Hu\Pk'''+\left[ {4\over 3} k^2 + 20
\left( \Hu'+2\Hu^2 \right) \right]\Pk'' + 6 \Hu k^2 \Pk'+{1\over
3}k^2 \left[ k^2 + 4\left( \Hu'+2\Hu^2 \right) \right]\Pk =0,
\label{Phi4}
\end{equation} 
\end{widetext}
\noindent
whose analysis indicates that the solution~(\ref{phi2}) is valid for
$| x|\ll x_2$, up to $|x|\approx 1$, where the source term in
Eqs.~(\ref{Bardeen}) become relevant but $k$ is still completely
negligible.  We have to match this solution with the other relevant
solution in this region, namely, the bounce solution (\ref{Phi0}),
which have three arbitrary constants. As confirmed by the numerical
analysis below, we have to set $\tilde{C}=0$ because the function it
multiplies goes like $1/x^2$, which should appear, and dominate, in
Eq.~(\ref{phi2}) in a region where both solutions are valid. Hence, we
only need to determine $\tilde{B}$ and $\tilde{A}$. That is why we do
not need to calculate the evolution of $\dphi$ in order to determine
such constants; this also explains why the initial conditions we
assume for the scalar field perturbation are irrelevant for the final
power spectrum.

The solution~(\ref{Phi0}) will propagate the Bardeen potential to the
other side of the bounce, to the region where $x$ is of order one. As
we are in a region where $k$ is negligible, the point of matching will
be chosen to be $x=-N\ll -1$, where $N$ does not depend on $k$ but is
large\footnote{We consider $N$ large but not large enough to neglect
terms of order $N^5$ in the expansion of $f_1$ in Eq.~(\ref{Phi0}). If
we neglect such terms, we loose the effect of the bounce in the
evolution of the perturbations. Also, considering $N=1$, without
approximations, would not change our qualitative results and the power
spectrum.}.

The result of the matching reads
\begin{equation} 
\tilde{A} = A_1 - {8\eta_0\over 45 a_0^4 N^5}A_2,
\label{At}
\end{equation}
and
\begin{equation} 
\tilde{B} = -{\eta_0\over 3 a_0^4}A_2.
\label{Bt}
\end{equation}

At $x = N\gg 1$, on the other side of the bounce, these solutions must
be matched with a solution similar to Eq.~(\ref{phi2}), namely
\begin{equation} 
\Pk^> = C_1-C_2{\eta_0^4\over 3a_0^4\eta ^3},
\label{phi4}
\end{equation}
yielding
\begin{equation} 
C_1 = A_1 - {16\eta_0\over 45 a_0^4 N^5}A_2,
\label{c1}
\end{equation}
and $C_2 = A_2$. For the power spectrum, the important term in
Eq.~(\ref{phi4}) is the constant $C_1$: as we are now back to a
regular expanding universe, the other term is a decaying mode which
rapidly becomes negligible. In $C_1$, the dominant term when $k\ll 1$
is the one proportional to $A_2$ which goes as $k^{-5/2}$, while the
other is proportional to $\sqrt{k}$.  Hence, we get
\begin{equation}
k^3 |\Pk\,\,(-x_1)|^2 \propto k^3 |\tilde{A_2}|^2{\eta_0^2\over
a_0^8N^{10}}\approx {\lP^2 \over a_0^2\eta _0^4N^{10}} k^{-2}.
\label{sp}
\end{equation}
yielding a spectral index $\nS=-1$.

\begin{figure}[t]
\includegraphics*[width=8.5cm]{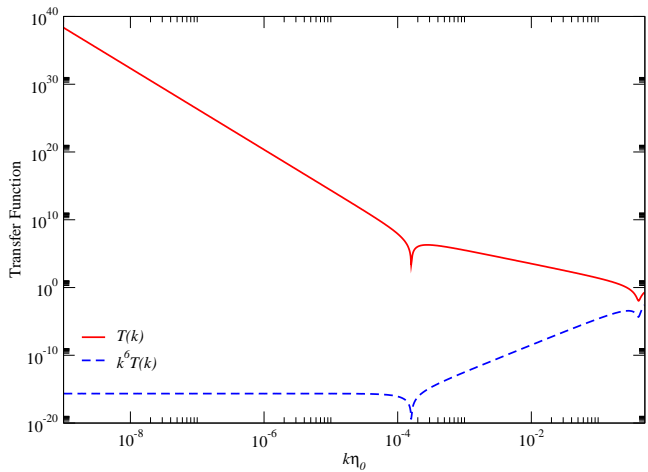}
\caption{Transfer function for the bounce model. Full line: ratio of
the squared gravitational potential amplitude between horizon exit and
re-entry. The dashed line shows the same multiplied by $\tilde k^6$ to
emphasize the power law behavior obtained in Eq.~(\ref{trans}).}
\label{fig:trans}
\end{figure}

One can then define a transfer function between ``Horizon exit'' and
``Horizon re-entry'' as the ratio of the power spectra at the
corresponding two different times. It is given in the case of our
bounce by the relation
\begin{equation}
T(k)={k^3 |\Pk\,\,(-x_1)|^2\over k^3 |\Pk\,\,(x_1)|^2} \propto (\eta _0
k)^{-6}.
\label{trans}
\end{equation}
This transfer function essentially depends on the behavior of the
scale factor at both times, as well as on the nature of the bounce
itself. It is represented on Fig.~\ref{fig:trans}

Let us now check all these approximations through a numerical
examination of Eqs.~(\ref{Bardeen}) and~(\ref{scalar}).

\subsection{Numerical calculations}
\label{sub:num}

The system~(\ref{Bardeen}-\ref{scalar}) can be solved numerically for
any value of $k$. For that purpose, we also include the characteristic
conformal timescale $\eta_0$ in the wavenumber $\tilde k=k\eta_0$ (and
correspondingly $\tilde \omega=\omega\eta_0$), and write the system as
\begin{equation} \left\{ \matrix{\displaystyle{\dd^2\Pk\over \dd x^2}
+ \displaystyle{4 x\over x^2+1} \displaystyle{\dd\Pk\over \dd x}
+\tilde \omega^2 \Pk &=& -\displaystyle{\sqrt{2} \over x^2+1}
\displaystyle{\dd X_k\over \dd x}, \cr \hfill\cr
\displaystyle{\dd^2X_k\over \dd x^2} + \displaystyle{2 x\over x^2+1}
\displaystyle{\dd X_k\over \dd x} +\tilde k^2 X_k &=&
\displaystyle{4\sqrt{2} \over x^2+1} \displaystyle{\dd \Pk\over \dd
x}, }\right. \label{dimensionless} \end{equation} relations in which
$X_k\equiv\lP\dphi$ (recall that $x\equiv \eta/\eta_0$), subject to
initial conditions, far in the limit $x\to -\infty$, given by
Eqs.~(\ref{phii}) and~(\ref{scai}) with $\etai=\etaj=0$ , namely
\begin{equation} 
\Pk^{^{\rm ini}} = -{3^{3/4}\alpha \over 2 x^2 \tilde k^{3/2}}
\biggl({\sqrt{3}\over \tilde k x}+i\biggr)\de^{-i\tilde k x/\sqrt{3}},
\label{condphinum}
\end{equation}
and
\begin{equation} 
X_k^{^{\rm ini}} = {\alpha\over x\sqrt{2\tilde k}}
\de^{-i\tilde k x},
\label{condXnum}
\end{equation} 
where we have defined the only free dimensionless parameter
$\alpha=\lP\sqrt{\eta_0}/a_0$. In all the figures, this parameter has
been arbitrarily fixed to the value $\alpha=10^{-3}$; the conclusions
do not, however, depend on this value, which acts as a simple
normalization constant.

The solution of Eqs.~(\ref{dimensionless}) for the square of the
Bardeen potential $|\Pk\,\,|^2$ is shown on the bottom panel of
FIG.~\ref{fig:spec} for various values of the wavenumber, renormalized
with the bounce characteristic conformal timescale, $\tilde k$,
ranging from $10^{-6}$ to $\sim 1$ on the figure as a function of the
renormalized conformal time $y\equiv \tilde k x = k\eta$. All
calculations are started far in the radiation dominated epoch, for
$y=-100$, where the boundary conditions hold. This is verified as,
indeed, for small enough values of $k\eta_0$, $|\Pk\,\,|^2$ behaves as
$\eta^{-4}$, as expected. It can be checked that, as discussed above,
in the long wavelength limit, the Bardeen potential starts with a
negligible constant part and a growing, $\propto \eta^{-3}$, mode, for
$-x_1\ll x\ll 1$, which then connects to the $f_1$ part while crossing
the bounce, and then connects back to the usual growing and decaying
modes, although the new constant part has now acquired a piece from
both modes of the previous epoch.

\begin{widetext}
Once the system~(\ref{dimensionless}) is solved, one can easily
compute the value of the Bardeen potential at horizon crossing, namely
for $x\sim 1/\tilde k$, i.e., $\eta\sim 1/k$, or $y\sim 1$. This
provides the spectrum shown on the top panel of
FIG.~\ref{fig:spec}. It is clear on that figure that for small values
of $\tilde k$, the behavior of the power spectrum is indeed a power
law, which we checked indeed corresponds to $\nS=-1$. Also shown is a
comparison between various cases of interest, namely the vacuum case
for which the initial conditions given by Eqs.~(\ref{condphinum})
and~(\ref{condXnum}) hold, the gravitational vacuum case for which
Eq.~(\ref{condphinum}) still holds, but with Eq.~(\ref{condXnum})
replaced by $X_k^{^{\rm ini}} =0$, and finally the decoupled case for
which the coupling between $\Pk\,\,$ and $X_k$ is made to vanish,
i.e. for which the left-hand side of Eqs.~(\ref{dimensionless}) is
arbitrarily set to zero. The curves corresponding to either vacuum or
gravitational vacuum initial conditions are seen to be almost
undistinguishable, showing that, as expected and discussed in the
previous section, the final spectrum for the gravitational potential
does not depend on the initial conditions for the scalar field
perturbations. The decoupled curve shows that, for $\tilde k\ll 1$, if
one were to neglect the bounce duration and apply some matching
conditions by brute force, one gets the same spectral index $\nS=-1$,
but with a normalization that is wrong by many orders of
magnitude. The situation is even worse for intermediate scales for
which even the index is wrong.

\begin{figure}[h]
\includegraphics*[width=14cm]{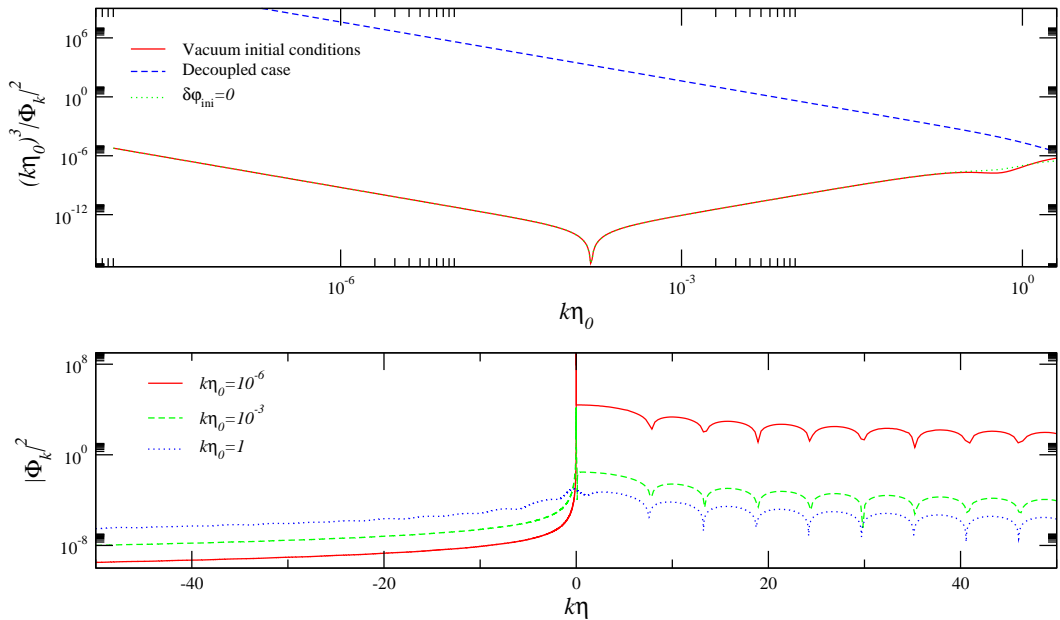}
\caption{Top panel: spectrum of scalar perturbations, i.e., $k^3
|\Pk\,\,|^2$ as function of the wavenumber $k$, normalized with $\eta_0$
as indicated. The long wavelength part of the spectrum, as expected,
is well fitted by a power-law with spectral index $\nS=-1$. The full
line is for vacuum initial condition for $\delta\varphi$, the dotted
line is with $\delta\varphi=\delta\varphi'=0$ at the initial time
($y_{\rm ini}\equiv k\eta_{\rm ini}=-100$ in the numerical
calculation), and the dashed curve represents the fully decoupled
situation for which $\varphi'$ is assumed negligible all along.
Bottom panel: Time evolution of the gravitational potential $|\Pk\,\,|^2$
for different wavelengths.} \label{fig:spec}
\end{figure}
\end{widetext}

\begin{figure}[h]
\includegraphics*[width=8.5cm]{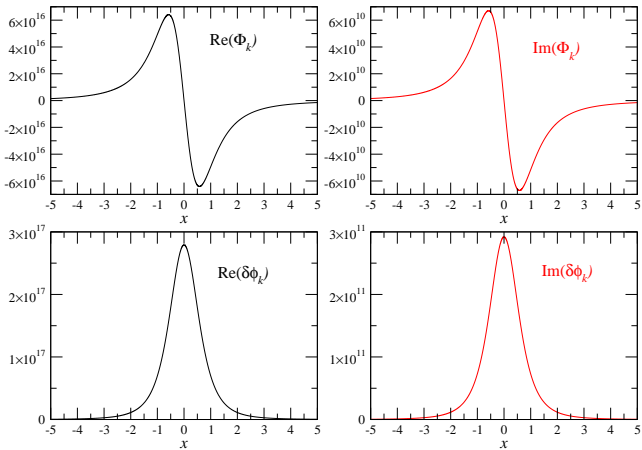}
\caption{Real (left) and imaginary (right) parts of the Bardeen
potential (top) and the scalar field perturbation (bottom) as a
function of $x=\eta/\eta_0$ for $\tilde k=10^{-8}$.}
\label{fig:reimsol}
\end{figure}

On FIG.~\ref{fig:reimsol} is shown an enhancement of the region
surrounding the bounce itself. This figure shows that the real and
imaginary parts of both the Bardeen potential and the scalar field
perturbation connect, respectively, with the bounce functions $f_1$
and $f_3$, thereby confirming the prediction $\tilde C=0$.

\begin{figure}[t]
\includegraphics*[width=8.5cm]{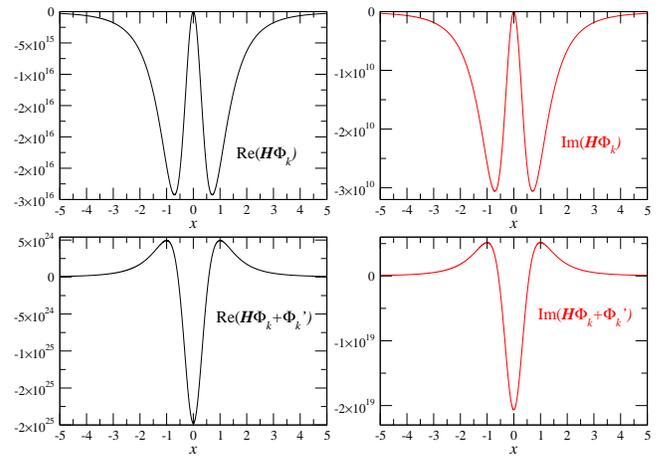}
\caption{Real (left) and imaginary (right) parts of the geometrical
junction variables $\Hu\Pk$ (top) and $\Hu\Pk+\Pk'$, (bottom) as a
function of $x=\eta/\eta_0$ for $\tilde k=10^{-8}$.}
\label{fig:junct}
\end{figure}

The fact that the Bardeen potential only connects to the odd bounce
function $f_1(x)$ suggests that in the limit in which the bounce
duration $\eta_0$ can be neglected, one may apply the following
junction conditions across the surface at which the bounce is
located ($\eta=0$)
\begin{equation} \left[ \Hu\Phi \right]_\pm = \left[ \Phi'+\Hu\Phi
\right]_\pm =0,\label{junc} \end{equation} where $\left[ A\right]_\pm
= A(-\etaB)-A(+\etaB)$ is the jump in the geometric quantity $A$. On
FIG.~\ref{fig:junct} are shown the time evolution across the bounce of
the quantities involved in Eq.~(\ref{junc}). For a fixed surface
thickness $\etaB$, and in a way independent of this thickness, the
relevant quantities are indeed conserved and can be safely used. This
is of course true only in the particular example presented here, but
it can also be conjectured to apply for a symmetric bounce in general.
In fact, taking the ekpyrotic model~\cite{ekp} with our matching
conditions~(\ref{junc}), one obtains a scale invariant spectrum.

\section{Conclusions}
\label{sec:conclusion}

We have presented a cosmological model in which a bounce takes place
in the framework of pure general relativity. This is achieved by
assuming that, at some stage after a contracting phase, a negative
energy free scalar field became important. Performing a bounce with
such a scalar field, instead of an ordinary hydrodynamical fluid,
permits to regularize the perturbation, which otherwise grow unbounded
near the bounce~\cite{nobounce}. We derived the last horizon crossing
spectrum and obtained, both analytically and numerically, a spectral
index $\nS=-1$ in the long wavelength limit, therefore ruling out such
a model as a competitor to the inflationary paradigm. However, our
study of a concrete bouncing model allowed us to obtain some intuition
on what happens with perturbations when they pass through a bounce.
First of all, the bounce acts indeed as a ``pump field'' for
perturbations. Secondly, the field which produces the bounce in the
background solution, and its perturbations, is not relevant for the
evolution of the Bardeen potential in almost the whole history of the
model, except near the bounce itself, where it becomes very important
for the power spectrum amplitude, although not for the spectral
index. Finally, usual matching conditions are not valid for
transitions through a bounce.  In fact, even for the background
metric, such conditions are not valid since the Hubble parameter
$H=\Hu/a$ changes sign through the bounce, by definition. Through our
bounce, the Bardeen potential also changes sign, and what happens to
be continuous is the combination $\Hu\Phi$. Inspired by our concrete
model, we suggested matching conditions to be applied to general
models where the bounce is not specified, which are spelled out in the
subsection \ref{sub:num}. Of course, these suggestions must be checked
within other concrete examples, or through a more general formal
analysis.

The model we have discussed is admittedly over simplistic. We may be
confident in its latest part describing the radiation dominated epoch,
which we know have taken place in our Universe, and accept that it may
provide a reasonable description of an immediately preceding bouncing
phase. There is, however, no reason to believe, even in the case of a
bounce, that the evolution of the Universe should have been symmetric
in time. One can instead set up a contracting phase with a different
scale factor, assuming at some stage some form of entropy production,
to end up with enough radiation before the bounce, yielding a scale
factor of the form~(\ref{scale}) that ultimately connects back to
standard cosmology. Such a model should originate in a realistic
underlying particle physics theory.

Let us briefly discuss an example, which is reminiscent of the ekpyrotic
proposal~\cite{ekp}, with a few differences. First, the model we have
in mind would be purely four dimensional and does not intend to
address the flatness problem. Second, such a model would be
effectively singularity free. Finally, we would not need to impose
arbitrary~\cite{mpps} matching conditions across the bounce to obtain
the required mixing in the growing and decaying modes before and after
the bounce, since its specific form would be known.

\begin{figure}[t]
\includegraphics*[width=8.5cm]{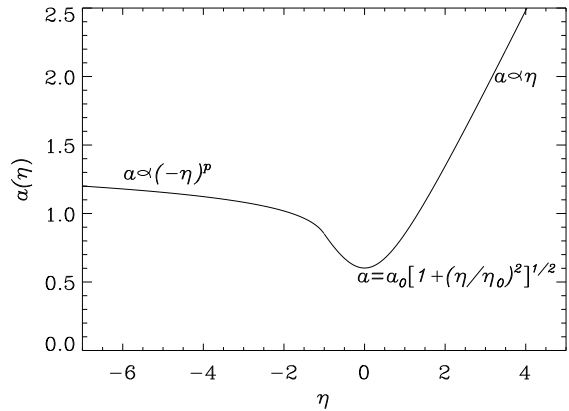}
\caption{Scale factor for connecting a slowly contracting phase to our
model. If such a four dimensional model was effectively constructed,
it would produce a scale invariant spectrum of perturbations and would
thus become a promising competitor to more usual inflationary models.}
\label{fig:abounce}
\end{figure}

More precisely, a model satisfying the abovementioned requirements
could consist in a bouncing model having a slowly contracting phase,
$a\propto (-\eta)^p$, with $0<p\ll 1$, connected to the phase examined
in the previous sections, as the one shown on FIG.~\ref{fig:abounce}.
When the perturbation in the Bardeen potential crosses the horizon for
the first time, the dependence $\eta^{-2}$ in Eq.~(\ref{phii}),
stemming from the fact that the universe is supposed to be radiation
dominated at that time, would be substituted by a dependence
$\eta^{-2p}$, i.e., almost independent of $k$ when $k\eta \sim
1$. Doing calculations along the lines of those presented in
Sec.~\ref{sec:pert}, the scale invariant spectrum follows.

\acknowledgments

We would like to thank CNPq of Brazil for financial support.  NPN
should like to acknowledge IAP for hospitality during the time this
work was being done. We also would like to thank Ruth Durrer, Gilles
Esposito-Far\`ese, Jean-Philippe Uzan and the group of ``Pequeno
Semin\'ario'' for various enlightening discussions. We are also
especially indebted to J\'er\^ome Martin for numerous important
comments and discussions.

\end{document}